# A *SwarmESB* Based Architecture for an European Healthcare Insurance System in Compliance with GDPR

Cristina Georgiana Calancea[1], Lenuţa Alboaie[2] and Andrei Panu[3]

[1] Faculty of Computer Science, Alexandru Ioan Cuza University of Iasi, Romania
`cristina.calancea@info.uaic.ro`
[2] Faculty of Computer Science, Alexandru Ioan Cuza University of Iasi, Romania
`adria@info.uaic.ro`
[3] Faculty of Computer Science, Alexandru Ioan Cuza University of Iasi, Romania
`andrei.panu@info.uaic.ro`

**Abstract.** With the everlasting development of technology and society, data privacy has proven to grow into a pressing issue. The bureaucratic state system seems to expand the number of personal documents required for any kind of request. Therefore, it becomes obvious that the number of people having access to information that should be private is on the rise as well. This paper offers an alternative cloud integration solution centered on user data privacy, its main purpose being to help software services providers and public institutions to comply with the General Data Protection Regulation. Throughout this proposal we describe how data confidentiality can be achieved by transitioning complex human procedures into a coordinated and decoupled swarm system, whose core lies within the "Privacy by Design" principles.

## 1 Introduction

Alongside with the evolution of society, people have become more prone to understanding how digitized systems could improve the everyday life experience. Moreover, the need for automation has surged from the countless irregularities regarding privacy discovered in the legal procedures that require public employees to go through citizen's personal data in order to validate it. The issue demanding a solution is the lack of coordination among governmental institutions and therefore, the failure to sustain the right to privacy.

General Data Protection Regulation [1], which took effect on May 2018, states that collecting, processing and storing user data without explicit consent is a punishable offense. In order to comply with regulations stipulated in the GDPR [1], many companies have sought ways to improve their systems as to support monitoring of the data access and information removal mechanisms. For instance, after the Cambridge Analytica scandal [2], Facebook is said to have improved its confidentiality policy by presenting the users choices regarding what they want to share and offering them the option to view and delete the data they store [4]. Even though GDPR is a regulation promoted by the EU, Google is also subjected to these laws when providing any of its services. Besides the explicit consent agreements, the enterprise has also worked on its ads mechanism by making targeting less aggressive [5].

Unlike private enterprises, governmental institutions find it difficult to cope with all the GDPR changes, since in most of them, the legal procedures are still executed by using pen and paper. A relevant example is Romania, an European country, where the health insurance system relies on excessive interaction between citizen personal data and human resources. The latter are necessary to complete almost any task, which often leads to privacy breaches. The obvious solution is the automation of the whole process, as a distributed system, which offers the same services, whilst shielding user privacy. According to the "Privacy by Design" principles [6], our digital processing unit must be built in order to foresee data breaches and integrate confidentiality in its components from the beginning, not as a last minute extension. This paper promotes an alternative cloud integration technology, Swarm ESB [7], which centers on protecting data confidentiality, while decoupling complex systems in small entities that can coordinate themselves in order to work with as little user data as possible.

## 2 Swarm Architecture as a Viable Alternative for Integration and Privacy

There are a few Enterprise Service Bus implementations that offer Integration Platform as a Service solutions, oriented on privacy: Mule ESB [8], WSO2 ESB [9] and Swarm ESB [7]. Mule ESB offers resource access constrained by several filters and policies, while preventing sensible data exposure by using encryption, digital signatures and access control techniques for APIs usage. WSO2 ESB integration solution implements sixteen security scenarios inspired from the web services security policies. Some examples are the Integrity, Confidentiality and Kerberos Token-based Security scenarios [12]. Each one of them uses either digital certificates or keys to verify the identity of the sender and the authenticity of the message.

In order to understand the way Swarm ESB approaches privacy and how it is modeled in Healthfuse (described in section 3), we will briefly present the concepts behind it - for a detailed perspective on SwarmESB, see [7] [10]. Swarm communication is a pattern of sending and processing messages between adapters. An adapter is a server side software node that offers a functionality of the system, which can be used only through a swarm. Usually, communication in an ESB implementation takes place between complex entities that process simple messages. Swarm ESB pictures messages as "smart" entities, capable of taking over some of the workload, by being routed between specialized components, which helps to reduce the complexity of distributed systems, offering scalability, availability, decoupling and parallel use of resources [10].

The integration strategy proposed by Swarm ESB is one of the few that cover all privacy principles, by introducing the usage of executable choreographies. The implementation of this concept turns formal contracts between organizations in code executed by every communication participant. The standard classification that helps us model various processes of integration and privacy assurance contains three categories, according to [11]. A privacy advantage of using verified choreographies is the capability of monitoring the data stream directly in the integration layer, which is logically separated by the processing code, found in adapters. Encrypted Choreographies use various control access mechanisms, with the purpose of identifying and authenticating the key entities that communicate through swarms, while serverless choreographies are appropriate for deployment in a public cloud, which offers monitoring and full automation of processes' capabilities. Executable choreographies are the key principle of the "Privacy-Integration" model Swarm ESB proposes.

## 3 Healthfuse - Swarm Based Architecture in Compliance with GDPR

### 3.1 Romanian Health Insurance System

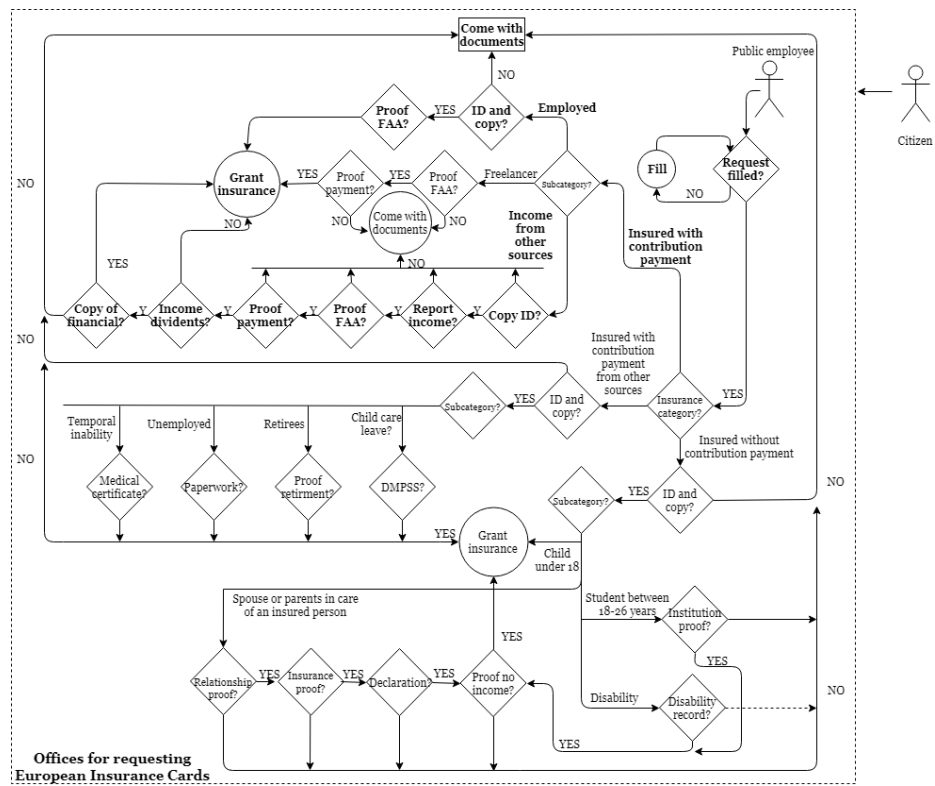

**Fig. 1.** Flow of interaction when requesting an European Health Insurance

The outline in Figure 1 summarizes the tedious experiment of obtaining a European Insurance Health Card for a person who is an employee and also has a business. The six steps that must be performed in order to obtain an insurance are: bringing the identity card and a copy of it, supplying an employment proof, providing an income declaration and its confirmation from the National Fiscal Administration Agency and showing the receipt which proves all taxes were paid. Having an extra income earned from dividends results in the need of bringing additional financial statistics from the Ministry of Finance. The total amount of time spent by going through all this flow revolves around ten hours, since we had to visit three public institutions in order to gather private paperwork and show it to at least five different people with the purpose of fulfilling procedures, while completely dismissing citizen's privacy.

### 3.2    Healthfuse - an Alternative Approach

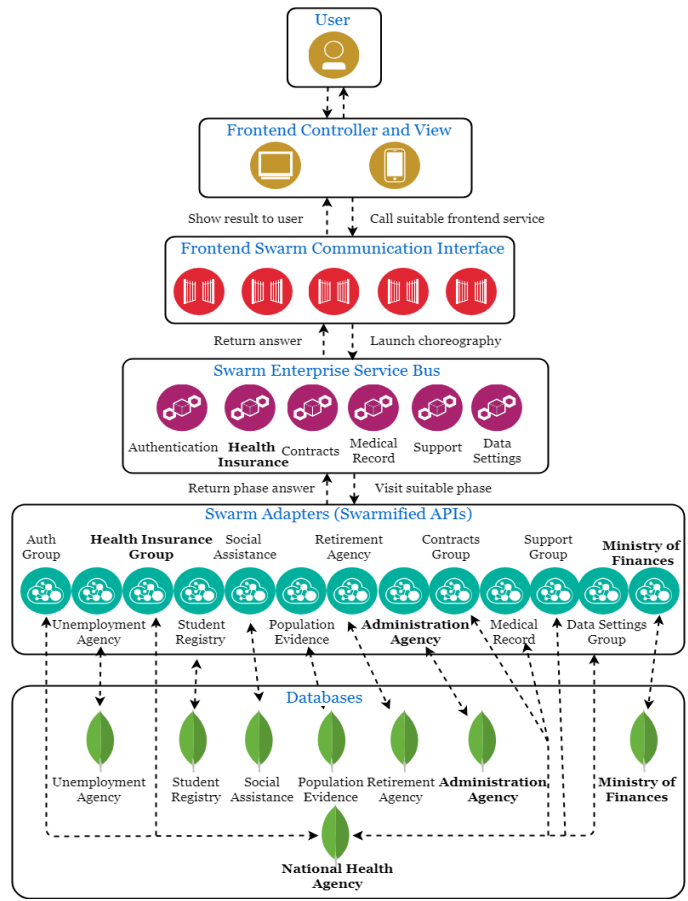

**Fig. 2.** HealthFuse Swarm Application Architecture

In order to turn privacy into a key player, the flow shown in Figure 1 is automated to reduce human interaction with user data, as it is outlined in the proposed Healthfuse architecture, described in Figure 2. It reduces the time consuming six steps process to a three steps flow: log in, select the type of insurance and then wait for the request to be validated by a swarm that executes the automated version of the process presented in Figure 1, synthesized in choreography. Since the suggested architecture is a distributed one, it enhances the benefits of using a cloud system to coordinate all entities and keep their stored data confidentially. Therefore, our alternative approach provides the assurance that any shared data is kept private and used only for automated processing requested by the user. The only scenario that leads to data disclosure to the technical team is when an inquiry is submitted by the user directly through the integrated support system.

Since the significant amount of information needed to validate the insurance flows cannot be found in the resources belonging to the National Health Insurance Agency, we need to create additional adapters used to extract data from entities like the Administration Agency, described on the fourth level of Fig. 2. Adapters supply data to swarms - outlined on the third level of the architecture - in order to transport it between entities, following various choreographies. This process spares citizens from going between public institutions to gather their personal data and from handing it to strangers. The architecture is designed by separating entities and allowing them to exchange data on authenticated channels only when necessary. That way, the relevant data is distributed to the relevant entity and accessible to the others only on valid premises. The main features Healthfuse offers to its users are the ability of getting the appropriate insurance policy from just a few easy steps and opening a medical record by uploading, deleting and downloading files they may need to have closed. In Table 1, we describe the way our proposal is designed and implemented in respect to the "Privacy by Design" Principles.

**Table 1.** Implementation of "Privacy by Design" Principles in Healthfuse

| Privacy by Design Principles | Healthfuse - A SwarmESB Based Solution |
|---|---|
| 1. Proactive not reactive; Preventative not remedial | Our solution offers an access monitoring mechanism, which makes it formally verifiable, since it can accurately state which entities has accessed the private data. |
| 2. Privacy as the default setting<br>3. Privacy embedded into design | Privacy concerns are treated by using verified choreographies in the swarm's implementation. This way, adapters only process data, leaving the confidentiality checks to the integration layer, composed by swarms. |
| 4. Full functionality; Positive - sum, not zero-sum | The business model can be altered at any time, without producing any impact over the current components and flows; instead we would just develop choreography to encapsulate the new model. |

| 5.Visibility and transparency – keep it open | The implemented choreographies target communication between many organizations, all of them being known to the others. Private data can be accessed only by entities that have the privileges to run the proper choreography. |
|---|---|
| 6. End-to-end security – full lifecycle protection | Data is protected by the SCRAM-SHA1 encryption protocol applied to the authentication step of the storage unit. End-to-end security is guaranteed by using communication through WSS Protocol on server side and HTTPS Protocol on client side. |
| 7. Respect for user privacy – keep it user-centric | Access to private data is done only with the consent of the user when reporting an incident in the support system. The user is prioritized over commercial interest by offering the option to delete any collected data. |

## 4    Conclusion

This paper presents the impact GDPR has on public institutions, the difficult situation in which they found themselves while trying to comply with the new requirements regarding data privacy and a solution to most of their problems, based on Swarm architecture.

Throughout this paper, we propose a way to stop the ascending tendency of losing the control of our personal data to various governmental agencies. Since the key principle in creating Healthfuse was to protect data, we remodeled the health insurance system from its roots and built it as an independent and specialized entity that only exchanges data with other software entities built on the same principles.

As for the future, this project can be looked at as a continuously developing system [13], which implies collaboration from several other public institutions like the Financial Administration Agency and Retirement Agency. The current prototype can be developed into a public administration tool which works relying on controlled collected data distribution, avoiding at the same time its centralization. This leads to the usage of a coordinated system which doesn't keep the data it uses secret.

Revolutionizing the healthcare insurance procedures by installing such a system would most likely reduce data leaks and would most certainly offer a better user experience in terms of time and resources.

**Acknowledgements**

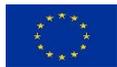 The dissemination of this work is partly funded by the European Union's Horizon 2020 research and innovation programme under grant agreement No 692178. It is also partially supported by the Private Sky Project, under the POC-A1-A1.2.3-G-2015 Programme (Grant Agreement no. P 40 371).